\documentclass[twocolumn,prb,showpacs,superscriptaddress,showkeys,preprintnumbers,amsmath,amssymb]{revtex4}

\usepackage{graphicx}
\usepackage{dcolumn}
\usepackage{bm}
\usepackage{multirow}
\usepackage{subscript}
\usepackage{nicefrac}

\begin{document}


\textheight 24.0 true cm

\title{Local magnetic anisotropy in BaFe$_2$As$_2$: a polarized inelastic neutron scattering study}

\author{N. Qureshi}

\email{qureshi@ph2.uni-koeln.de} \affiliation{$II$. Physikalisches
Institut, Universit\"{a}t zu K\"{o}ln, Z\"{u}lpicher Strasse 77,
D-50937 K\"{o}ln, Germany}

\author{P. Steffens}

\affiliation{Institut Max von Laue-Paul Langevin, 6 rue Jules
Horowitz, BP 156, 38042 Grenoble Cedex 9, France}

\author{S. Wurmehl}

\affiliation{Leibniz-Institut f\"{u}r Festk\"{o}rper- und
Werkstoffforschung (IFW) Dresden, D-01171 Dresden, Germany}

\author{S. Aswartham}
\affiliation{Leibniz-Institut f\"{u}r Festk\"{o}rper- und
Werkstoffforschung (IFW) Dresden, D-01171 Dresden, Germany}

\author{B. B\"{u}chner}

\affiliation{Leibniz-Institut f\"{u}r Festk\"{o}rper- und
Werkstoffforschung (IFW) Dresden, D-01171 Dresden, Germany}

\author{M. Braden}

\email{braden@ph2.uni-koeln.de}

\affiliation{$II$. Physikalisches Institut, Universit\"{a}t zu
K\"{o}ln, Z\"{u}lpicher Strasse 77, D-50937 K\"{o}ln, Germany}

\date{\today}

\begin{abstract}

The anisotropy of the magnetic excitations in BaFe$_2$As$_2$ was
studied by polarized inelastic neutron scattering which allows one
to separate the components of the magnetic response. Despite the
in-plane orientation of the static ordered moment we find the
in-plane polarized magnons to exhibit a larger gap than the
out-of-plane polarized ones indicating very strong single-ion
anisotropy within the layers. It costs more energy to rotate a
spin within the orthorhombic {\it a-b} plane than rotating it
perpendicular to the FeAs layers.

\end{abstract}


\maketitle

The recent discovery of the iron-based oxypnictide
superconductors~\cite{1} has triggered enormous activity exploring
the superconducting pairing mechanism in these materials. Density
functional theory (DFT) calculations~\cite{2,3} indicate that the
conventional electron-phonon coupling is by far too weak in the
iron pnictides to account for the observed high critical
temperatures of up to $T_c=55 K$~(Ref.~\onlinecite{4}). Instead,
the phase diagrams of many FeAs-based families suggest a magnetic
pairing mechanism, as superconductivity appears close to an
antiferromagnetic spin-density wave (SDW) phase.\cite{5,6,10}
Superconductivity can even coexist with the SDW order parameter
which then slightly decreases upon entering the superconducting
state.\cite{10a,10b,10c} Furthermore, the onset of
superconductivity in doped BaFe$_2$As$_2$ compounds is accompanied
by the appearance of additional magnetic scattering at the SDW
wave vector.\cite{7,8,9}

\begin{figure}
\includegraphics[width=0.41\textwidth]{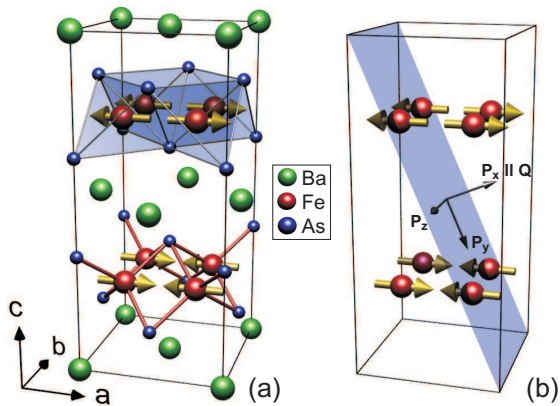}
\caption{\label{fig:structure} (Color online) (a) Visualization of
the crystal and magnetic structure of BaFe$_2$As$_2$ in the
orthorhombic setting. (b) Definition of the reference frame for
the neutron polarization analysis on the (1 0 1) reflection as an
example (only the Fe ions are depicted in the unit cell).}
\end{figure}

BaFe$_2$As$_2$ becomes superconducting on hole-doping,\cite{10} on
electron-doping~\cite{11} or by the application of
pressure.\cite{12,13} Pure BaFe$_2$As$_2$ exhibits a structural
phase transition from the high-temperature tetragonal ($I4/mmm$)
to the low-temperature orthorhombic phase ($Fmmm$), see
Fig.~\ref{fig:structure}(a), at $T_S\sim$140~K
(Ref.~\onlinecite{14}) accompanied by a magnetic phase transition
into a collinear antiferromagnetic structure with an ordered Fe
moment of 0.87~$\mu$\textsubscript{B} (Ref.~\onlinecite{14}). This
orthorhombic antiferromagnetic phase exhibits remarkable
properties. The magnetic structure shown in Fig. 1(b) breaks
tetragonal symmetry, one may therefore expect some orthorhombic
splitting in the crystal structure as it indeed occurs, but this
splitting remains rather small. However, the signature of the
orthorhombic structure in the electronic properties is much
stronger than the small orthorhombic splitting in the lattice
constants would suggest. Resistivity measurements on detwinned
single crystals reveal pronounced anisotropy associated already
with the precursors of the orthorhombic phase.\cite{16} The
observation of a larger resistivity along the shorter axis with
ferromagnetic spin coupling is further surprising. Also scanning
tunnelling microscopy \cite{17} and optical conductivity analyses
\cite{18} report essential in-plane anisotropy. Magnon excitations
in the SDW phase have been studied by several inelastic neutron
scattering experiments (unpolarized) and, again, the slope of the
low energy dispersion is anisotropic.\cite{19,19a,20,20a} The
nearest neighbor interaction along the $a$ and $b$ directions
needed to model the magnon dispersion even show opposite signs.
This remarkable anisotropy in magnetic exchange, however, can
qualitatively be explained by DFT.\cite{21} The pronounced
anisotropies in the electronic properties inspired a model of
orbital polarization \cite{orbital} which gets support from recent
ARPES experiments,\cite{22,23} as well as speculations about the
role of nematic properties.
\newline In this work we analyze the magnetic single-ion anisotropy
which can be deduced from the gap in the magnetic excitations
finding again remarkably strong in-plane anisotropy. It costs more
energy to rotate the spin within the FeAs layers than aligning it
perpendicular to them, in contrast to the simple expectation of an
easy-plane system.

Large BaFe$_2$As$_2$ single crystals were obtained by a self-flux
technique.\cite{24} The pre-reacted precursor materials FeAs,
Fe$_2$As, Co$_2$As and BaAs were mixed leading to a Ba(Fe,
Co)$_{3.1}$As$_{3.1}$ composition. This composition was used to
achieve a homogeneous melt at T = 1463~K. The melt was cooled
slowly under a temperature gradient in a double-wall crucible
assembly to obtain large and flux-free single crystals of
BaFe$_2$As$_2$. Using this technique, single crystals with lateral
dimensions up to 25 x 10 mm$^2$ and thickness up to 1 mm were
obtained.  Two plate-like single crystals with masses of 804 mg
and 200 mg have been coaligned in order to decrease counting time.
The experiment has been carried out at the thermal-beam three-axis
spectrometer IN20 (ILL) using polarizing Heusler (111) crystals as
monochromator and as analyzer. Longitudinal polarization analysis
was performed with a set of Helmholtz coils to guide and orient
the neutron polarization.  The longitudinal polarization analysis
permits separating the nuclear cross-section and the individual
components of the magnetic cross-section into the spin-flip (SF)
and non-spin-flip (NSF) channels by selecting the initial and the
final polarization direction of the neutrons. We use the
conventional reference frame for polarization analysis,\cite{25}
see Fig.~\ref{fig:structure}(b), with the $x$ axis parallel to the
scattering vector $\mathbf{Q}$, $y$ perpendicular to $x$ within
the scattering plane, and $z$ perpendicular to $x$ and $y$
(perpendicular to the scattering plane). The sample has been
mounted in the [100]/[001] scattering geometry in orthorhombic
notation, but note that the crystal is twinned so that part of the
crystal is in [010]/[001] geometry and signals from both parts
superpose.

Longitudinal polarization analysis adds additional selection rules
to the general law in neutron scattering that only magnetic
components perpendicular to the scattering vector contribute. Only
the magnetic components perpendicular to the axis of polarization
analysis contribute to the SF scattering whereas the parallel
components contribute to the NSF scattering.\cite{25} Nuclear
scattering is always a NSF process. We measured the NSF
cross-section for polarization parallel to $x$, serving as a
reference for eventual contaminations, and all three SF
cross-sections keeping $\mathbf{k}_f$ constant. Applying a
correction for the finite flipping ratio of $R=7.4$ the magnetic
cross-sections $\sigma_y$ and $\sigma_z$ (Eqs.~\ref{eq:sigmay}
and~\ref{eq:sigmaz}) as well as the background in the SF channel
(Eq.~\ref{eq:with}) can be calculated according to:
\begin{equation}
\sigma_y={R+1\over R-1}[I(SF_x)-I(SF_y)]), \label{eq:sigmay}
\end{equation}
\begin{equation}
\sigma_z={R+1\over R-1}[I(SF_x)-I(SF_z)], \label{eq:sigmaz}
\end{equation}
\begin{equation}
BGR_{SF}(R)=\frac{R}{R-1}[I(SF_y)+I(SF_z)]-\frac{R+1}{R-1}I(SF_x).
\label{eq:with}
\end{equation}

\begin{figure}
\includegraphics[width=0.45\textwidth]{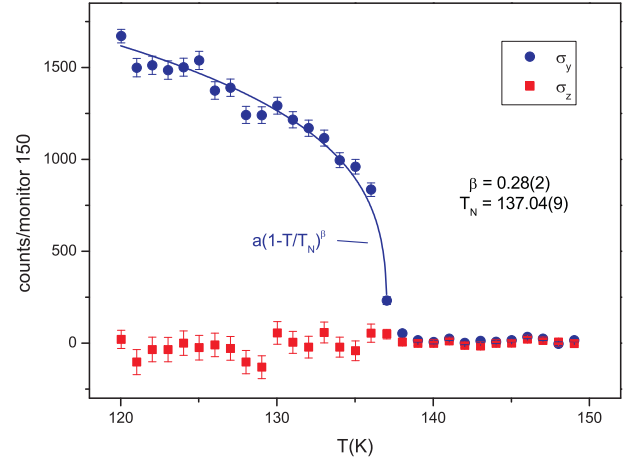}
\caption{\label{fig:T_N} (Color online) Temperature dependence of
the elastic magnetic scattering cross section at $\mathbf{Q}$=(1 0 1) revealing that almost the entire magnetic signal
stems from $\sigma_y$. The values have been obtained
by Eqs.~\ref{eq:sigmay}
and~\ref{eq:sigmaz}. The fitting of a power law to the $\sigma_y$ data yields a N\'eel temperature of 137 K.}
\end{figure}

In Fig.~\ref{fig:T_N} the maximum intensity of the magnetic
\mbox{(1 0 1)} reflection is shown as a function of temperature
for the two channels. A power law has been fitted to the $\sigma
_y$ data revealing a $T_N$ of approximately 137 K. The entire
elastic signal is found in $\sigma_y$ ($\sigma_z$ is essentially
zero) in perfect agreement with the well-accepted magnetic
structure shown in Fig.~\ref{fig:structure} with moments parallel
to the $a$ axis.

\begin{figure}
\includegraphics[width=0.47\textwidth]{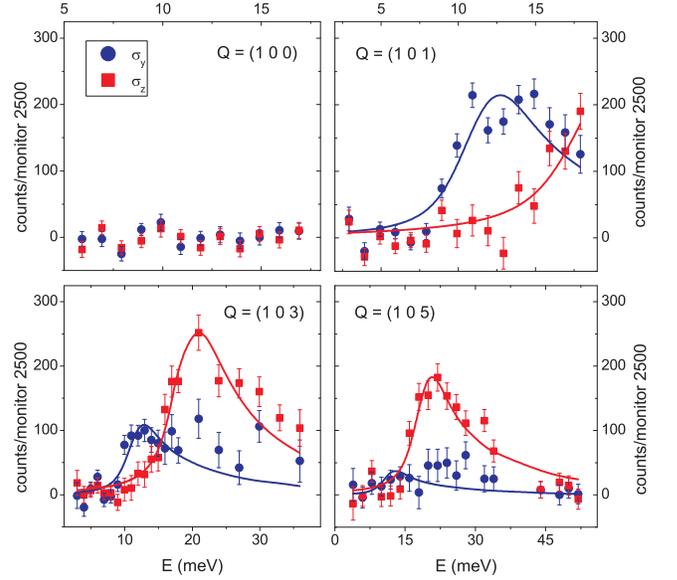}
\caption{\label{fig:inel} (Color online) Inelastic
constant-$\mathbf{Q}$ scans at the magnetic Brillouin-zone centers
(1 0 $l$) for different $l$ values and at (1 0 0). (Blue) dots
represent the magnetic scattering cross section $\sigma_y$, while
(red) squares depict $\sigma_z$. The cross sections have been
obtained by measuring the SF$_z$ and SF$_y$ channels,
respectively, and applying Eqs.~\ref{eq:sigmay}
and~\ref{eq:sigmaz}. The solid lines are spline-interpolated
spin-wave theory calculations folded with the experimental
resolution.}
\end{figure}

Inelastic constant $\mathbf{Q}$-scans were performed at the
magnetic zone centers $\mathbf{Q}$=(1 0 $l$) with $l$=1,3 and~5,
and at $\mathbf{Q}$=(1 0 0) which is a magnetic zone boundary due
to the finite interlayer coupling (Fig.~\ref{fig:inel}). The scan
at \mbox{$\mathbf{Q}$=(1 0 0)} does not indicate any magnetic
scattering in agreement with the published magnon dispersion for
CaFe$_2$As$_2$ (Refs.~\onlinecite{19,19a}) and for SrFe$_2$As$_2$
(Refs.~\onlinecite{20,20a}) indicating energies of the order of
100 meV for this $\mathbf{q}$ value. Due to the high zone-boundary
energies the magnetic signals of the differently oriented twin
domains do not superimpose in the scans at $\mathbf{Q}$=(1 0 $l$)
with $l$=1,3 and~5, which are zone boundaries for the second
domain orientation. The magnetic-zone-center scans sense the gap
of the magnon dispersion arising from anisotropy. For a system
with uniaxial anisotropy one expects a single gap for the two
still degenerate transversally polarized magnons. This degeneracy,
however, lifts for orthorhombic local symmetry causing an easy, a
medium and a hard axis.\cite{26} We will see that such
orthorhombic single-ion anisotropy is needed to describe magnetic
excitations in BaFe$_2$As$_2$. The cross sections $\sigma_y$ and
$\sigma_z$ unambiguously show the splitting of the zone-center
magnons in BaFe$_2$As$_2$. The $z$ direction of our reference
system is always parallel to the $b$ direction and therefore
$\sigma_z$ measures the transverse magnon polarized along the $b$
direction. The $y$ direction lies always in the
$a^\star$,$c^\star$ plane but rotates with varying $Q_l$. Since
the ordered moment is parallel to $a$, $\sigma_y$ contains only
the transverse magnon polarized along the $c$ direction (see below
for discussion of the contribution of the longitudinal magnon).
Due to the geometry this signal is reduced by a factor 0.92, 0.61,
and 0.42 for  $Q_l$=1,3, and 5, respectively. The qualitative
analysis of the scans at the magnon zone centers immediately shows
that the in-plane magnon gap is larger than the out-of-plane one.
This is remarkable in view of the in-plane orientation of the
ordered moment.\cite{14} Starting from the tetragonal symmetry at
high temperature one might describe BaFe$_2$As$_2$ as an
easy-plane system similar to many layered magnets with
K$_2$NiF$_4$ (214) structure, e.g. K$_2$FeF$_4$
(Ref.~\onlinecite{27}). In such a system the two transverse
magnons still split as a higher-order cubic in-plane anisotropy
adds to the dominating easy-plane anisotropy. There is no
mass-less Goldstone mode in such system. But in this case the
energy of the in-plane polarized transverse magnon is below that
of the out-of-plane one.\cite{27} The fact that BaFe$_2$As$_2$
exhibits the opposite behavior, the out-of-plane mode lies around
10\ meV below the in-plane mode near 20\ meV, excludes the
description of the magnetic phase in an adapted tetragonal model,
which, however, works very well for many 214-materials. The local
single-ion anisotropy of the SDW phase in BaFe$_2$As$_2$ is truly
orthorhombic.\cite{26} This observation is in line with the strong
electronic signatures \cite{16,17,18} in this phase which contrast
with the only weak orthorhombic splitting of the lattice. The
strong single-ion anisotropy within the FeAs-layers gives strong
support for pronounced orbital polarization. Previous unpolarized
neutron scattering experiments on SrFe$_2$As$_2$, CaFe$_2$As$_2$
and BaFe$_2$As$_2$ (Ref.~\onlinecite{27a}) most likely reported
only on the lower gap as the different polarizations could not be
resolved.

In order to quantitatively analyze the scattering distribution we
have folded the experimental resolution with the scattering
function of the magnon contribution using the Reslib
code.\cite{28} The dispersion and the structure factors of the
magnons of the AFe$_2$As$_2$ compounds have been calculated in
linear spin-wave theory by Yao and Carlson \cite{30} for the
Hamiltonian including nearest-neighbor exchange coupling along $a$
and $b$ directions, $J_a$ and $J_b$, next-nearest neighbor
exchange, $J_2$, inter-layer coupling, $J_c$, and anisotropy
terms. A gap at the magnetic zone center appears only for a finite
single-ion anisotropy term $\sum_i{-\Lambda (S_i^z)^2}$ where, in
our case, $\Lambda$ is equal\cite{note} to $\Lambda_b$ or
$\Lambda_c$ and $S_i^z$ is the component of the spin operator of
spin i in the direction of the ordered moment. Yao and Carlson
obtain for the gap at the magnetic zone center: $\Delta =
2S\sqrt{\Lambda \cdot(2J_a+4J_2+\Lambda +2J_c )}$. For the
exchange parameters we have used the values recently determined
for SrFe$_2$As$_2$ (Ref.~\onlinecite{20}) but normalized them to
the value $S$=${1\over 2} \cdot 0.87$ corresponding to the ordered
moment in BaFe$_2$As$_2$, note that the full shape of the
dispersion is not essential for our analysis as we only sense the
low-energy part of it in our measurement. With $S$=${1\over 2}
\cdot 0.87$, $SJ_{1a}$=30.8 meV, $SJ_{1b}$={\mbox -5 meV},
$SJ_2$=21.7 meV, $SJ_c$=2.3 meV, $\Lambda_b$=0.99(2) meV, and
$\Lambda_c$=0.38(1) meV, we may simultaneously describe the two
channels measured at the three zone centers, with the anisotropy
terms and the scale factors as the only free parameters. This
description takes into account the $Q$-dependence of the magnetic
scattering due to the form factor and the geometrical factor given
above. The calculated values have been interpolated with a
B-spline in order to obtain continuous curves which are depicted
as solid lines in Fig.~\ref{fig:inel}. The perfect description of
the experimental data validates the assumption that the total
magnetic scattering arises from the transversally polarized
magnons. Using the obtained anisotropy values we determine the
in-plane and the out-of-plane gap to amount to 16.4 meV and 10.1\
meV, respectively.

The single-ion anisotropy parameters obtained in our experiment
can be compared to the result of relativistic DFT calculations
including spin-orbit coupling;\cite{31} the authors find that only
0.16 meV/Fe are necessary to turn the magnetic moment from the
easy axis to the $b$ direction, but that 0.20 meV/Fe are required
to turn the moment perpendicular to the plane. These calculations
cannot properly describe the local single-ion anisotropy in
BaFe$_2$As$_2$ as the sign of the anisotropy does not agree with
the experimental values. DFT \cite{31} thus misses something in
the description of the local symmetry of the SDW phase in
BaFe$_2$As$_2$.

\begin{figure}
\includegraphics[width=0.45\textwidth]{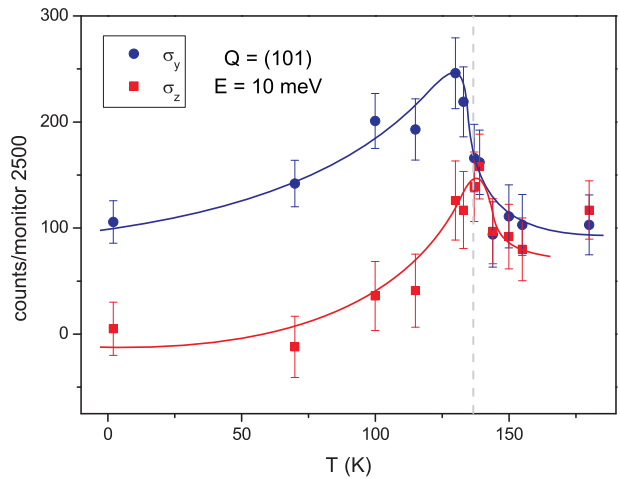}
\caption{\label{fig:ineltemp} (Color online) Temperature
dependence of the inelastic magnetic scattering at $\mathbf{Q}$=(1
0 1) and an energy transfer of 10 meV documenting the transition
from isotropic to anisotropic fluctuations at $T_N$. The solid
lines serve as a guide to the eye.}
\end{figure}

In an itinerant system with non-saturated ordered moments, one may
expect a longitudinal magnetic excitation, i.e. a magnon polarized
along the ordered moment. This mode contributes to the $\sigma_y$
channel, but by comparing the different $\mathbf{Q}$-values
studied we can fully rule out such an interpretation for the
studied energy ranges. The longitudinal contribution should
increase with increasing $l$ component, but the experimental data
show the opposite and are well described by the geometry factors
for the transverse mode. We can thus excude a longitudinal
excitation below $\sim$35\ meV in agreement with DFT studies
predicting them at significantly higher energies.\cite{32,33}

The temperature dependent measurement of the two magnetic cross
sections at an energy transfer of 10\ meV illustrates how the
isotropic magnetic fluctuations in the paramagnetic phase
transform into the anisotropic magnons below the N\'eel
temperature (Fig.~\ref{fig:ineltemp}). There is no difference
visible between the two channels at high temperature. Upon
approaching the magnetic transition the intensity increases due to
the softening of the magnetic response. Below the transition, the
magnetic signal is almost completely suppressed in the $\sigma_z$
channel where the larger gap develops, but there is only moderate
reduction in the $\sigma_y$ channel, where the energy transfer of
10\ meV is just slightly below the low-temperature maximum.

In conclusion the polarized inelastic neutron scattering studies
on the magnetic excitations in BaFe$_2$As$_2$ reveal a strong
in-plane single-ion anisotropy which contrasts with the small
structural orthorhombic distortion in this material. This
pronounced local anisotropy corroborates the strong electronic
signatures of the orthorhombic phase and gives further support for
an important role of orbital degrees of freedom in the iron
pnictides. The local anisotropy of Fe in the SDW phase apparently
is insufficiently described in DFT theory even when including
spin-orbit coupling.

\begin{acknowledgments}
This work was supported by the Deutsche Forschungsgemeinschaft
through the Sonderforschungsbereich 608, through the
Forschergruppe 538 (grant (BU887/4)), as well as through grant
BE1749/12 and BE1749/13. S. Wurmehl acknowledges support by DFG
under the Emmy-Noether program (Grant No. WU595/3-1). We thank  M.
Deutschmann, S. Pichl, K. Leger and S. Gass  for technical support
and A. Yaresko for the discussion about the DFT calculations.

\end{acknowledgments}


\begin{thebibliography}{}
\bibitem{note} The calculation in reference \cite{30} was done for uniaxial
anisotropy, but the case of orthorhombic anisotropy calculated
with only nearest-neighbor exchange \cite{26} shows that the two
magnon gaps arising from orthorhombic single-ion anisotropy are
identical to the two gaps caculated with these parameters in
uniaxial anisotropy.
\end{thebibliography}
\end{document}